# High-Pressure Synthesized Materials: a Chest of Treasure and Hints

V.V. Brazhkin

*Institute for High Pressure Physics,*
*142190 Troitsk, Moscow region, Russia*

The present review covers the production of new materials under high pressures. A primary limitation on the use of pressures higher than 1 GPa is a small volume and mass of a produced material. Therefore, despite an extremely wide range of new high-pressure synthesized substances with unique properties, synthesis on an commercial scale is applied up to now only to obtain superhard materials, this real treasure of today's industry. At the same time, high-pressure experiments often give material scientists a hint at what new intriguing substances can exist in principle. This is true for new superhard, semiconducting, magnetic, superconducting and optical materials already synthesized under pressure, and as well as for a large number of hypothetic new polymers from low-Z elements. Many of new materials, including the above polymers, may exist in the metastable form at normal pressure at sufficiently high temperatures.

**Introduction**

The synthesis of materials under high pressure is a vast area of physics, chemistry and engineering. Should you browse the Internet to search "high pressure synthesis", the web search engines, for example, Google, would respond with over 10 mln links, making it absolutely impossible to write any comprehensive review on this topic. For this same reason, neither it is possible to list all relevant references. The purpose of this paper is to present the author's personal view on the state of art and prospects of the production of new materials by using high-pressure synthesis, and to complement already existing reviews, for example [1,2]. It should be noted that a number of publications, for example [2], discuss a comparatively narrow range of materials synthesized under pressure. Other publications, for example [1], also focus on anomalous properties of materials in a strongly compressed state, although most of materials with such properties cannot be retained at normal pressure, hence, they are not directly related to the subject under consideration. In addition, the behavior of materials in a strongly compressed state has also been extensively covered in scientific literature (for example, see [3]).

For physicists, the history of a high-pressure research is first of all associated with P.W. Bridgman, a 1946 Nobel Prize Laureate "for the invention of an apparatus to produce extremely high pressures, and for the discoveries he made therewith in the field of high pressure physics". Nowadays high-pressure synthesis of new materials is primarily associated with a successful synthesis of superhard materials - diamond and cubic boron nitride - in the second half of the $20^{th}$ century. At the same time, the pressures $1 – 10^3$ MPa have been already in use for more than a century in chemistry for the industrial production of materials. What's more, the development of relevant engineering procedures has been more than once rewarded with a Nobel Prize in chemistry. In 1918, F. Haber won the Nobel Prize "for the synthesis of ammonia from its elements…"; in 1931 C. Bosch and F. Bergius shared the Nobel Prize "in recognition of their contributions to the invention and development of chemical high pressure methods…".

Chemical industry, biotechnology, crystal growth procedures now widely employ autoclave presses operating at pressures of $1 – 200$ MPa. But such pressures are regarded as being rather low in view of the present-day level of high-pressure technologies, and I will leave this area aside in the subsequent discussion. The more so as comprehensive reviews and studies on corresponding problems are available (for example, see [4]). Technological applications involving a combined effect of pressure and shear stresses, for instance, in hydro-extrusion, also have been covered in detail in special literature. Powder metallurgy and high pressure sintering of ceramics and alloys do not enter into the scope of this paper either. In the present review, I will be concerned with the use of pressures higher than 1 GPa to prepare new materials. It should be mentioned that over 90% of the visible substance in the Universe are subject to the aforesaid pressures, since the substance is primarily concentrated in large self-gravitating bodies – planets and stars. Let me also remind here that the pressure in the ocean depth is $10 – 10^2$ MPa; in the center of the Earth it amounts to ~340 GPa.

## 1. Matter under Pressure: from Physics to Biology

Let us consider the meaning of the term "high pressures" on the basis of the degree of pressure effect on the substance. In physics, there is a natural quantity $P_0$ with pressure dimension

$$P_0 \sim Ry/a_b^3 \sim m^4 e^{10}/h^8 \sim 10 \text{TPa}$$

where $Ry$ is the rydberg, $a_b$ is the Bohr radius, $m$ is the electron mass, $e$ is the electron charge, $h$ is the Planck's constant.

Physically, the value $P_0$ corresponds to the fact that the action of pressure on the electrons in a substance becomes comparable to the internal interaction of the electrons with the ions. For $P \gg P_0$, periodic dependence of the substance' properties on the atomic number changes to a weak monotone dependence. In this case all substances become "dull" like metals. Thus, we can conditionally correlate the energy effect scale with the pressure scale. For characteristic atom and ion sizes, the pressure of 1 TPa corresponds to the energy of 1 – 10 eV; 1 GPa corresponds to the energy of 10 – 100 K (1eV ~ $10^4$ K), although the effect of pressure is of course not equivalent to that of temperature. Various substance modifications (phases) differ in energy from one another by the values within the range 0.01 – 1 eV. Therefore, the majority of phase transitions in simple substances fall in the range 1 GPa – 1 TPa. This is the working range of high pressures for physicists - material engineers. Similarly we can estimate the "essential" pressures for chemistry ~ 0.1 GPa – 10 GPa, and for biology ~ 0.01 GPa – 1 GPa, etc. (see Fig 1).

The degree of the effect of pressure on the substance also is strongly dependent on the type of the interparticle interaction. The above estimate of the pressure range for high-pressure physics refers to the substances with a strong metallic, covalent or ionic bonding. For the substances with a weak intermolecular interaction (Van-der-Valse forces), the pressures 0.1 – 10 GPa are substantially high because the atom size is considerably larger. In this connection, it is sometimes more convenient to speak not about pressure but about the degree of compression of a substance. For the substances with a strong interatomic bonding, the bulk modulus is B ~ 100 GPa, and for molecular substances and rare gas solids B ~ 1 – 10 GPa. As a result, the degree of compression by tens of percent in the given classes of substances is reached, respectively, at different pressures of the order of the bulk modulus values. That is, for molecular substances, the pressures, at which the structure and properties of a substance can be significantly modified, amount to 1 – 10 GPa, while the figures for most metals and covalent and ionic substances are 10 – $10^3$ GPa.

## 2. High Pressure Apparatus: the Ways to Compress

Despite the radical action on the substance, the use of high pressures is not as widespread in material chemistry as, for example, the use of high temperatures and catalysts. The reason for this lies, firstly, in the complexity and high cost of high pressure equipment, and, secondly, in the smallness of volumes in which high pressure can be obtained. The advancement of high pressure physics, chemistry, and material engineering is to a great extent dictated by the origination of new high pressure equipment, including high pressure apparatus, press machinery, compressors, etc. Static pressures of 0.1 – 1 GPa can be created in fairly large volumes ~0.01 – 1 $m^3$ in gas containers, autoclave presses, and high pressure apparatus of a piston-cylinder type. The creation of pressures higher than 1 GPa requires the use of a high pressure apparatus with moving parts. For more than a century, there have been used two, in a certain sense, alternative types of high-pressure devices, those of a piston-cylinder and Bridgman-anvils type [5] (see Fig. 2a). The volumes of a high pressure zone in a piston-cylinder device are 1 – 1000 $cm^3$, in a number of cases they reach ~1 $m^3$ (very large presses are needed in this case (see Fig. 2d)), but maximum pressures amount to 0.5 – 3 GPa, and it is only some unique small-volume piston-cylinder apparatus that have allowed reaching pressures of 5 – 7 GPa. Much higher pressures can be produced in the apparatus of a Bridgman-anvils type: for hard alloy anvils – 15–20 GPa; for SiC anvils – 20-70 GPa; for diamond anvils – 100-300 GPa. However, the volume of a high-pressure zone in these devices is very small and ranges, depending on pressure, from $10^{-6}$ to $10^{-1}$ $mm^3$. A possible future use of large synthetic diamonds or SiC crystals can hardly affect the situation considerably, for the strength of the anvils inevitably decreases with their sizes [6]. The need for obtaining pressures over 5 GPa in relatively large volumes ~1$cm^3$ for the synthesis of superhard materials – diamond and cubic boron nitride (c-BN) – has given rise to the development of new high pressure devices incorporating the ideas contained in the piston-cylinder and Bridgman-anvils designs [7]. Most-used and widely recognized types of high-pressure apparatus have become the "belt", multianvil apparatus, "toroid", etc. (see Fig.2a-c). All these devices can be employed for the creation of pressures 3 – 8 GPa in the volumes 1 – 10 cm$^3$; pressures 8 – 15 GPa can be obtained in the volumes 0.1 – 1 cm$^3$. Pressures 15 – 50 GPa in relatively large volumes ~1 mm$^3$ can be obtained if using double-stage versions of the above devices.

In addition, there are methods for the creation of dynamic pressures mainly by the shock-wave technique that easily permits reaching pressures ~100 – 1000 GPa. However, short times of dynamic pressure action (nanoseconds) and extremely high temperatures accompanying the pressure pulse severely restrict the employment of dynamic high pressures for the synthesis of new materials at pressures P > 100 GPa. Nevertheless, dynamic pressure methods are used for producing materials in large volumes 1 – 10 cm$^3$ at P ~10-100 GPa, in particular, for the synthesis of nanodiamonds.

Small volumes (1 mm$^3$ – 1 cm$^3$) and sizes of the specimen of new materials obtained at pressures over 1 - 10 GPa set extremely severe limits on the application of the high-pressure technique for material chemistry. The industry of superhard materials is almost the only and "lucky" exception, as it appears to be "satisfied" with its superhard products of a few mm in size. Various new attractive ultrahard, semiconducting, superconducting, magnetic, optical and other materials obtained under pressure are unlikely to be ever produced by the high-pressure technique commercially. In this connection, speaking of a "renaissance of high pressure material chemistry" [1] is somewhat hastily. At the same time, high pressure experiments, in this instant, give a hint at what new interesting materials are expected to be created in principle. Quite frequently, many of these materials can be obtained at normal pressure through non-equilibrium material chemistry techniques, such as chemical vapor deposition (CVD), plasma sputtering, chemical reaction, the use of new catalysts, etc. In so doing, the restrictions on the volumes or/ and sizes of obtained materials are removed. The examples of such hints from high pressure experiments to material science will be considered below in Sections 6,7.

## 3. Obtain and Retain

Unfortunately, many substances existing at high pressure cannot be retained at room pressure. For instance, perovskite-structure (Mg (Ca), Fe)SiO$_3$ minerals are most abundant substances on Earth but they have never been observed at normal pressure due to the instability during decompression [8]. Striking examples of new alkali and alkali-earth metal modifications with weird structures [9]; metallic and superconducting phases of alkali-halids (CsI), chalcogenes (O, S), rare gas solids (Xe), metalloids (B), normally magnetic metals (Fe) [10]; polymerized nitrogen [11]; hypothetical metallic hydrogen [12]; high pressure phases of melts and glasses [13] are no doubt in the forefront of the high pressure science itself. However, they do not have a direct relationship to the high-pressure material synthesis because these materials either can never be recovered to normal pressure conditions or can be maintained at room pressure only at ultra-low temperatures. Hence, it does not seem appropriate to consider these particular accomplishments in this review.

In most papers on high-pressure chemistry, a lively discussion of a possible metastability or instability of substances produced under high pressures inevitably ends with something like "on condition that the substance can be conserved". At the same time, the possibility of conserving a high-pressure material can be analyzed in advance [14]. Of course, it does not concern pressure-synthesized **stable** substances like ammonia NH$_3$. Here we deal with phases that do not match the Gibbs' free energy minimum, that is, with metastable phases. The notion of the "phase" of a substance was introduced by J.W. Gibbs [15] as a state of substance with a certain "phase", meaning a set of particles with definite pulses and coordinates that make up this substance. The name "metastable phase" is given to a non-equilibrium state of a substance whose properties do not change or change reversibly over a period of an experiment or observation. Strictly speaking, there are no metastable phases in classical thermodynamics, since through infinite time the system should irreversibly relax to the equilibrium state with the lowest value of the Gibbs' free energy. However, there are long-lived metastable solid phases whose life-time at normal conditions exceeds the time scale of the Universe, and their existence cannot be neglected. A number of crystalline metastable phases, such as diamond, have the P,T-region of thermodynamic stability. Other metastable modifications like amorphous solids or fullerite $C_{60}$ have no stability region at any pressure and temperature. Some metastable substances like silicon clathrates seem to have conditional stability region at negative pressures.

The metastable condition of phases is provided for by the existence of the energy barrier on the way of the transformation of a system into a low-laying energy state (see Fig.3). At reaching certain temperatures or/and pressures, metastable phases often relax through energy-intermediate states to a thermodynamically equilibrium ground state. Thus, amorphous materials upon heating crystallize; diamond, if heated at normal pressure, transforms to graphite, and so on.

In general, the problem of the stability of metastable phases including those prepared under high pressure should be analyzed on the basis of a microscopic approach. The stability of a phase means that eigen vibration frequencies of its lattice are positive. In the region of the elastic or vibration instability the lattice cannot exist for times exceeding $\sim 10^{-13}$ s and undergoes a phase transition, which prevents the recovery of high-pressure modifications (see Fig.4). However, the *ab-initio* calculation of the phonon spectrum at finite temperatures is very difficult to be made, and predictions about the regions of the kinetic temperature stability for particular metastable high-pressure phases are very few in number. If at positive pressures a high-pressure phase loses its elastic or dynamic stability even at zero temperature, it can never be conserved at room pressure at any temperatures (see Fig.4 (curve (a)).

Certain important conclusions on the stability of high-pressure phases can be drawn on the basis of simple thermodynamic arguments. A metastable high-pressure phase has, by definition, the Gibbs' free energy G higher than the one for a stable phase. As a result, the temperature of thermodynamic equilibrium between the metastable crystal and stable melt, $T_m^*$, is always lower than the equilibrium melting temperature for the stable phase, $T_m$ (see Fig.5). It seems that $T_m$ is not only a thermodynamic boundary for the existence of a metastable high-pressure phase but also the upper limit for their kinetic stability. Indeed, the nucleation of the stable melt at the crystal surface occurs without activation for both stable and metastable crystalline phases due to the flat geometry of nucleation, a small surface tension associated with surface reconstruction of the crystalline lattice, and to the absence of elastic stresses because of the zero static-shear modulus of the liquid phase. The rate of the melting process, occurring without the activation barrier at the surface of a solid, will be constrained in the volume by the relaxation processes determined by melt viscosity.

In practice all metastable high-pressure phases on heating at normal pressure transform to more stable ones at far lower temperatures than the melting temperature of a stable phase $T_m$. For example, diamond on heating transforms to graphite at $T\sim 1500$ K, which is about one third of the melting temperature. The exception is the substances with close $T_m^*$ and $T_m$ values and very viscous melts, for instance, $B_2O_3$, for which the high-pressure phase melts into a stable melt without intermediate crystallization into a stable phase [16].

Thus, high-pressure modifications of low melting temperature molecular substances like $H_2$, $O_2$, $N_2$, etc., either cannot hold the metastable form at normal pressure at all or can exist only at ultra-low temperatures. It is interesting, however, that many polymerized modifications obtained under pressure from molecular substances can exist at normal pressure in the metastable form at fairly high temperatures. The reason for this resides in the fact that the initial molecular substances are metastable in themselves, that is, they do not meet the Gibbs' free energy minimum. An excellent example of such transformation is pressure polymerization of molecular ethylene with the formation of polyethylene [17]. This question will be discussed in greater detail in Sections 4 and 7.

## 4. New Substances and New Properties

High pressure can modify in different ways the structure and properties of substances, resulting in synthesis of new materials. Here we make an attempt at systematizing possible ways of producing new materials by high-pressure methods (see Fig.6). Up to now the systematization of high-pressure synthesis has been carried out primarily according to the classes of materials obtained (superhard substances, electronic, optical materials, etc.). If classifying by the method of the production of new substances, it would be worthwhile to consider separately the stable and metastable pristine materials (see Fig.6).

The most widespread technique of obtaining new substances at high pressure consists in the synthesis of an equilibrium high-pressure phase that is maintained after the release of pressure. Typical examples of such a phase are diamond made from graphite and coesite and stishovite made from quartz. In doing so, both low- and high-pressure phase have their associated thermodynamic stability regions in the P,T-diagram of the substance. In the case of multi-component substances, the type of a phase diagram can vary under pressure; for example, new compounds, equilibrium under pressure, can emerge, while elementary substances at normal pressure do not form any compounds. Thus, alkaline and alkali-earth metals do not form compounds at normal pressure with 3d-metals. If compressed, they do form pressure-stable intermetallic compounds with the given metals [2,18].

In a number of instances, especially at moderate temperatures, transformations under pressure proceed into energy-intermediate "kinetic" phases with their associated sufficiently low activation barrier for the transformation. In the case of the graphite-diamond transition, such "intermediate" phase is lonsdaleite (hexagonal diamond). Glasses, obtained by quenching from melt at high pressure [19], as well as amorphous phases obtained by solid state amorphization [20] also can be classed with "kinetic" non-equilibrium phases obtained by means of high pressure.

Moreover, the pressure that affects transformation kinetics, including nucleation and the grain growth rate, can modify micro- and macrostructure of the material at the nano- and meso-level, specifically, grain size and morphology, texture, defect structure and concentration, and so on. In the process, by varying P,T-conditions one can control morphology and structure of crystal grains both in a stable, at normal pressure, modification and in a high pressure phase. An excellent illustration of the possibility to control the micro- and macrostructure of various modifications of $GeO_2$ is provided in [21] (Fig.7). Solid state amorphization [20] also can be viewed as a limiting case of a displacive-like phase transformation accompanied by a large volume jump and occurring in the nano-scales [21].

Last but not least, high pressures in combination with high temperatures can influence the state of the optical defect centers in crystals [22] and glasses [23], modifying their optical properties, among them color. The applications of this pressure effect will be discussed below in Section 6.

In the event that the initial state of substances is metastable, high pressures can be utilized not only for the production of equilibrium high-pressure phases but also for the synthesis of stable normal-pressure phases, and, of course, of various intermediate-energy "kinetic" modifications (see Fig.6). Thus, from the mixture of $H_2$ and $N_2$ at high pressure one can easily synthesize a thermodynamically equilibrium compound - ammonia ($NH_3$). Metastable molecular white phosphorous at high pressure transforms into stable covalent black phosphorous. The decrease under pressure of the viscosity of melts of some oxides and chalcogenides allows obtaining crystals of stable modifications poorly crystallized at room pressure. Quartz, $B_2O_3$ and $As_2S_3$ are good cases in point.

There are also a great number of examples of the synthesis of new energy-intermediate kinetic phases from initially metastable modifications. For instance, fullerite $C_{60}$ at high pressures and temperatures transforms into equilibrium carbon phases of diamond and graphite through a number of intermediate polymerized structures and amorphous modifications [24].

Initial simple molecular phases in the C-O-N-H systems deserve a separate discussion. Because of the relative simplicity of these compounds, physicists are often unsuspicious of the fact that the majority of the condensed phases of these compounds are not thermodynamically equilibrium ground states. For instance, molecular ethylene $C_2H_4$ is not the ground state of the system, in particular, polyethylene $C_{2n}H_{4n}$ presents a more low-laying energy modification. The same is true for other hydrocarbons like acetylene $C_2H_2$, benzene $C_6H_6$, etc., since the only thermodynamically stable compound in the C-H system is methane $CH_4$. This can be readily seen from the data on the energy of a corresponding molecular formation [25], which is for $H_2$ – 430 kJmole$^{-1}$, $CH_4$ – 1642 kJmole$^{-1}$, $C_2H_4$ – 2225 kJmole$^{-1}$, $CH_2$ – 753 kJmole$^{-1}$, $C_2H_2$ – 1626 kJmole$^{-1}$; cohesive energy for graphite is 712 kJmole$^{-1}$; the energies of the formation of condensed phases of molecular substances do not exceed dozens of kJmole$^{-1}$. In a similar way in the C-O-H system, $H_2O$, $CH_4$ and $CO_2$ are the only ground states for their composition, as, for example, carbon monoxide CO, spirits or glycerole are actually metastable phases. At moderate pressures P ~ 1 GPa and temperatures T ~ 500 K all these metastable molecular compounds don't transform to the stable phases and they are in the quasi-equilibrium region of the P,T-diagram (see Fig.8). However, at sufficiently high pressures these molecular substances irreversibly transform into polymer modifications (Fig.8). As with polyethylene, the above polymeric phases at normal pressure can have fairly high temperature stability, much higher than the melting temperature of the initial molecular phase. Thus, the temperature stability of polyethylene is restricted by the graphite-methane liquidus temperature. In Section 7 we will return to the discussion of this subject.

Now from the classification by the method of producing new high pressure materials we pass on to a more common classification according to the classes of obtained substances (see Fig.9).

## 5. Superhard Materials: More Equal than Others

Superhard materials – synthetic diamond and cubic boron nitride – are major commercial high-pressure materials whose world annual production is over 200 ton. For many years, this outstanding achievement of high-pressure material science has been generating an immense scientific interest in superhard materials as such, and has been encouraging attempts to discover, through the high-pressure synthesis, new kinds of superhard substances. Because of a huge commercial importance of superhard materials, we discuss them in a separate section.

It is well known that macroscopic mechanical characteristics, including hardness, are governed not only by microscopic parameters of the material (the type of atom, structure and atomic forces) but also by the morphology of the substance of different space scales, defects in the sample, methods of measurement, temperature, and so on [26]. Nevertheless, there exists the natural upper limit of a possible hardness, that is, the "ideal" hardness [26], controlled by the elastic moduli of the substance (see Fig.10). In this respect, the strategic search and creation of new superhard materials should follow two main directions: first, the selection of materials with high elastic moduli, and, second, the modification of grain morphology and the state of defects to approach the "ideal" values of mechanical characteristics.

Elastic moduli of the material depend only on the interatomic interaction and atomic structure. For solids the well-understood correlation is observed between the moduli and effective electron valence density $\rho_{el} = N/V_m$, where N is the effective electron valence (the number of valent electrons per atom) and $V_m$ is the molar volume. This correlation corresponds to the dependences $B \sim \rho_{el}^\mu$, where $\mu$ - 1.25, and $G \sim \rho_{el}^\beta$, where $\beta$ ~1.47 [26] (see Fig.11a,b). The more rigorous correlation between G and $\rho_{el}$ is observed for the elements of a similar electron structure, the dependences vary between the $G \sim \rho_{el}$ and $G \sim \rho_{el}^{5/3}$ (see Fig.11b). A stronger dependence is observed for covalent substances. Summarizing, elastic moduli are mainly controlled by the electron valence density, and the search for new superhard materials should be conducted among substances with high values of the atomic and electron density and, accordingly, with high bonding energies. Therefore, it is reasonable that one should look for new superhard materials among dense high-pressure phases.

Superhard materials can tentatively be divided into three classes:

i) Covalent and ionic-covalent compounds from the middle of the $2^{nd}$ and $3^{rd}$ periods of the periodic table. These materials have large electron and atomic densities due to a small inner electron core size.

ii) Various carbon materials. Although carbon belongs to the elements of the $2^{nd}$ period, it is quite pertinent to consider carbon materials as a separate group. First, diamond and its hexagonal analog, lonsdaleite, are substances with the highest moduli (especially shear modulus) and highest hardness. Second, vast diversity of various crystalline and amorphous carbon materials, as well as predicted new carbon structures, makes it possible to regard carbon as a model element for the physics of superhard materials.

iii) Partially covalent compounds of transition metals and Pt-group metals with low-Z elements. Large coordination numbers in these compounds can be combined with a strong covalent interaction. These materials (for example, tungsten carbide, titanium diboride) are of great importance for industrial applications as, in addition to high hardness, they display a high fracture toughness.

In the first class, the most renowned material produced under pressure is cubic boron nitride, c-BN, ranked second after diamond in hardness and moduli and finding a great many commercial uses. Interestingly, the question of whether c-BN presents a high-pressure phase or has a thermodynamic stability region at normal pressure [27] is still open. A wurtzite-like c-BN phase is evidently a kinetic phase similar to lonsdaleite. Superhard boron carbides $B_4C$ and $B_{13}C_2$ are obtained at normal pressure. However, the B-C system is promising for the production of new high-pressure phases [28]. Another promising binary system for superhard high-pressure phases is the C-N system. Theory indicates to the existence of new superhard forms of the crystalline $C_3N_4$ and $C_{11}N_4$ compounds with elastic moduli close to those of diamond [1,29]. Under pressure, various amorphous superhard phases of $CN_x$ and a number of new polymeric and low-density "graphene" structures have been obtained thus far [1,29], although a successful synthesis of dense crystalline phases at ultra-high pressures ~50 GPa has been reported [30]. The $Si_3N_4$ compounds synthesized at normal pressure are in a wide commercial use. Recently a high-moduli and high-hardness spinel-structured form of $Si_3N_4$ has been obtained by the high pressure synthesis at P > 15 GPa [31]. The $B_6O$ phase seems to be an equilibrium room-pressure phase, but for producing large-size crystals possessing high moduli and hardness above 30 GPa one needs pressures of 3–7 GPa [1]. Incidentally, pure boron apparently has a superhard high-pressure phase too [26], though the pressures to obtain this phase are obviously quite high (>50 – 70 GPa). In this connection, a shock-wave synthesis of the new boron phase offers more promise. Among oxides of low-Z elements, it is worthy to mention a superhard oxide named "stishovite", a high-pressure phase of $SiO_2$ [32] (Fig.12). Recent high precision measurements of moduli and hardness carried out on the large single crystals of stishovite have given the following values: B = 315 GPa, G = 222 GPa, H for different directions - 31.8 GPa and 26.2 GPa [32]. Stishovite, apart from its hardness, is a very important material of the Earth interior. At the megabar range pressures, stishovite transforms into denser and probably harder phases with $PbO_2$- and pyrite-type structures [33]. An interesting example of a new covalent metastable high-pressure phase with high moduli is $CO_2$ [34]. At normal pressure and low temperatures $CO_2$ presents a soft molecular crystal. A hard covalent high-pressure phase of $CO_2$ seems to be recovered at the normal pressure – low temperature conditions.

The search for new high-pressure phases in the triple and quadruple systems of low-Z elements is still at the initial stage. A successful synthesis in the B-C-N system of the high-pressure phases $BC_2N$ and BCN with high moduli and hardness close to that of diamond is of particular interest [35]. These modifications obviously present "kinetic" high-pressure phases. There are reports on the successful synthesis of new triple high-pressure phases in the B-C-O system [36], as well as on the synthesis of the hard Si-Al-O-N oxynitride spinel [37]. Other superhard high-pressure triple and quadruple systems based on the B, C, N, O, Be, Al, Si, P compounds are yet to be discovered.

Let us now consider high-pressure carbon materials. Hypothetic ultrahigh-pressure carbon phases with the BC8 and R8 structures [38] are predicted to be denser than diamond. However, these phases have not been obtained yet because of the need for using multi-megabar range pressures, and, besides, the bulk moduli and hardness values for these phases are hardly superior to those for diamond [26,38].

A large family of new superhard carbon materials has been obtained by high-pressure treatment of initially highly metastable carbon modifications, such as fullerites $C_{60}$ and $C_{70}$, nunotubes, etc. [24,39]. For example, the heating of $C_{60}$ under high pressure of 1 – 15 GPa has permitted the synthesis of a number of polymeric, $sp^2 – sp^3$ – amorphous, and nanocrystalline diamond modifications [24] (Fig.13) having bulk moduli and hardness comparable to those of diamond and exceeding those of many other materials (see Fig.14a,b). The modifications in question are "kinetic" intermediate-energy carbon phases.

Among new III-group high-pressure phases, noteworthy are those of the $TiO_2$, $RuO_2$, $OsO_2$ and $ZrO2$ oxides with the high values of moduli and the hardness of 20 – 40 GPa [40] (Fig.15a,b). Examples of superhard high-pressure phases also are provided by new modifications of TaN and ReC (H~ 25 GPa and 35 GPa, respectively) [26]. The bulk moduli of these phases have not been measured. Recent reports have announced a successful synthesis of a new class of nitrides PtN, $Pt_2N$, IrN, etc. with sufficiently high values of the bulk moduli [41]. High moduli also should be common to a number of borides, for example, $OsB_2$.

The search for new high-pressure substances with high elastic moduli will undoubtedly be continued, especially in connection with the attainment of a megabar pressure range in a diamond anvils-type high-pressure apparatus and the use of a laser heating. The existence of compounds with compressibility lower than that of diamond seems possible, although the associated excess of their bulk moduli cannot be higher than tens of percent. In my opinion, diamond will remain a substance with the highest value of the shear modulus. The point is that the basis for high moduli is high atomic and electron density of substances, and one can hardly expect miracles of any of the high-pressure phases. In this context it should be noted that the experimental study of the high elastic moduli presents quite a difficult task. For the measurements to be reliable, we must have sample of a large size and, preferably, of a single crystal form. Unfortunately, in most cases the samples to be examined are small-sized, inhomogeneous, textured, etc. In the X-ray or neutron diffraction study of the bulk moduli under high pressure, hydrostaticity of a pressure- transmitting medium is a key factor. For example, the 3d-polymers of fullerite $C_{60}$ have the bulk moduli around 280 GPa [42], while measurements in the non-hydrostatic conditions give absolutely wrong values of approximately 600 GPa [43]. In the ultrasonic and Brillouin scattering studies, isotropy and homogeneity of a sample are prerequisites for correct measurements of the elastic moduli [26]. When scaling the moduli, you can be mistaken by 3-4 times if the samples are textured [44]. Hence, any data obtained for small-sized, poorly crystallized samples have to be considered invalid. For example, the statement that a diamond nanorods composite has the bulk modulus B ~ 490 GPa, which is 11% greater than the value for diamond [45], is obviously untrue. Undoubtedly erroneous is information on record high bulk moduli of carbon phases obtained from fullerite $C_{60}$: B ~1000 – 1600 GPa [46].

As our expectations with regard to ultrahigh bulk moduli are unlikely to be met, more promising appears the way of bringing hardness closer to the "ideal" magnitudes while remaining in the range of the same bulk modulus values. It should be mentioned that for the majority of hard materials the real and "ideal" hardness (or strength) differ by 3-10 times [26].

Hardness and strength of the material are known to increase and approach their "ideal" values in the two limiting cases:
i) for a defect-free single crystal virtually without dislocations (for example, a single-crystalline whisker), and

ii) for an amorphous or nano-crystalline state. In this case, the formation and motion of dislocations are hindered due to the defective structure of the material at the nanometer scale. The nanostructures of 5-20 nm grain-size are optimal for reaching the maximum hardness [47].

Both of these approaches have been successfully implemented in the synthesis of superhard diamond materials under pressure. Recently, there have been reports on the growth of large (0.2 carat) single crystal defect-free diamonds of the II-a type with nitrogen impurity below 0.5 ppm [48]. Such single crystals have the Knoop hardness of 145 GPa, which is by 1.5 times greater than the values for ordinary natural diamonds. These impurity-free defect-free diamonds have already found application as indenters for hardness testing machines [48].

Amorphous and nanocrystalline diamond materials have been successfully synthesized from graphite and soot [49], carbon nanotubes [39], fullerite $C_{60}$ [50]. The nano-size of the grains has been achieved through direct catalyst-free transformations of carbon materials into diamond with a low grain growth rate and a high nucleation frequency. Through the control of the transformation kinetics, materials with both round and lamel grain morphology have been produced [49]. Diamond nano-polycrystals have record hardness values (~ 130 – 150 GPa) and the fracture toughness coefficient ~10-15 $MNm^{-3/2}$, which is 1.5-2 times higher than the values for ordinary diamonds. Unfortunately, very high pressures of ~15-25 GPa necessary to obtain such diamond nanomaterials restrain their commercial production.

Nanomaterials based on c-BN have not been obtained thus far. However, high hardness values of materials in the B-C-N system [35] also seem to be associated with the nano-size of the grains for these substances.

Aside from the use of transformations at high pressure, nanomaterials can be obtained by way of compacting and sintering pristine nanoparticles. However, porosity, chemical admixtures, and stresses at the inter-granular boundaries hinder the production of materials with optimal mechanical characteristics. Thus, compacts fabricated by the sintering of diamond nanoparticles usually have the hardness of 20-40 GPa. A more advanced approach suggests the use of a high-pressure infiltration technique. The infiltration of diamond nanoparticles with the Si-melt at P~ 7 GPa results in the formation of a diamond SiC nanocomposite with the hardness of ~80 GPa [51]. There is a large area of the industrial production of diamond- and c-BN-based composite superhard materials but it falls outside the scope of the present review.

In addition, record hardness, strength and fracture toughness of nanomaterials offer a further technological advantage, namely, the possibility of attaining the highest quality of the surface polishing and sharpness of the cutting edge.

## 6. Pile of Hints

Apart from superhard materials, many high-pressure substances reveal intriguing and potentially beneficial characteristics, among them electron-transport, superconductive, magnetic, semiconducting, optical, thermal, dielectric and other properties. These new materials are described here in one section as most of them are obtained at pressures higher than 3 – 10 GPa and the volume of the obtained substances does not exceed cubic mm or cubic cm. As a result, these materials have not yet come into commercial use. Nevertheless, even in this case we are dealing with the technology of new materials, because it is just the high-pressure synthesis of new materials that gives researchers hints at the possibility of obtaining and retaining such substances in the metastable form at room pressure. Hence, one can seek for the synthesis of the given materials not through the high-pressure treatment but through the use of catalysts or by non-equilibrium methods (chemical reactions, plasma sputtering, chemical vapor deposition (CVD), rapid cooling from melt, etc.). In doing so, restrictions on the linear dimensions of samples, at least, in certain directions are lifted.

Among high-pressure synthesized materials, there are a great number of wide-gap semiconductors. Thus, GaN and its analogs AlN and InN, as well as ternary crystals like Ga(Al)-N make up a new class of wide-gap semiconductors highly advantageous for a variety of electronic applications, including the manufacture of diodes and lasers operating in the blue-green-ultraviolet optical regions. Despite the fact that these semiconductors with a wurtzite-like structure at normal pressure present equilibrium stable modifications, they can hardly be obtained as single crystals without using pressure because these materials melt incongruently, and growing crystals from their melts is impossible. Various high-pressure techniques for growing single crystals of these materials have been developed. For instance, they can be grown in a gasostatic container in the nitrogen atmosphere at high temperatures and pressures of 1-2 GPa [52], by cooling from the congruent melt at P- 6 GPa and T – 2500 K [53], and through hydrothermal synthesis from a relevant ammonia-rich solution in a piston-cylinder apparatus at P~ 2 GPa [54]. The last-named method is thought to be the most promising.

Allied materials, equally important for optoelectronics, in the Ga-O-N system (gallium oxynitride) with the spinel structure have been recently synthesized at high pressures 4-7 GPa both as polycrystals [55] and large single crystals of several mm in size [56].

Doped semiconducting diamonds and cubic boron nitride, for example, C:B, C:P, c-BN:Be, also are materials of much promise for optoelectronics. At present, semiconducting diamonds are chiefly obtained by the CVD process at low pressures, although single crystalline semiconducting diamonds of larger sizes are grown under pressure. With the semiconducting boron nitride (doped c-BN), the CVD method is not as good as with diamonds, and the high-pressure synthesis remains the only way of producing this material [57]. From the doped-c-BN on spontaneous p-n- transitions, light-emitting diodes were already manufactured [57].

The already mentioned modification $Si_3N_4$ with the spinel structure is not only a superhard material but also a wide-gap semiconductor with a direct gap [31]. Wide-gap semiconductors with a possibility of controlling the value of the gap include, as well, solid solutions $Si_3N_4$-$Ge_3N_4$ with the spinel structure and oxynitride spinels Si-Al-O-N [1,37].

An interesting material incorporating both high hardness and semiconductor properties is boron suboxide $B_6O$ mentioned earlier [1]. A hypothetical high-pressure phase of pure boron seems to be a metal or semimetal.

Many metastable carbon phases obtained under pressure from fullerites, nanotubes and carbines [58] are semiconductors with the gap value ranging from 0.1 eV to 3 eV, as distinct from graphite (a semimetal) and diamond (an insulator). The possibility of controlling semiconducting properties and mechanical characteristics by means of pressure through the change in the substance structure and atom fraction in sp-, $sp^2$- and $sp^3$-states is of great interest today [58].

The method of solid-phase amorphization (SSA) of high-pressure phases allows the production in the volume state of tetrahedral amorphous semiconductors based on Si and Ge and of $A^{III}B^{V}$ compounds [59]. At ambient pressure, the tetrahedral amorphous semiconductors can only be obtained as thin submicron films, but despite this fact they have many commercial uses, for example, as solar battery elements. Comparing physical properties of the bulk samples of pressure-synthesized amorphous diamond-like semiconductors with those of thin films can be very prompting for the development of new semiconductor technologies.

High pressures are also widely employed for the production of substances featuring interesting magnetic, thermo-electric and superconducting properties. For example, alkali and alkali-earth metals at normal pressure do not form compounds with transition metals due to a great difference in the atomic radii and electro-negativity. At high pressures, alkali and alkali-earth metals acquire the properties of d-metals and easily react with transition d-metals to form intermetallic compounds, a number of which is retained in the metastable state after the release of pressure. The first examples of the systems with new compounds were the K-Ni, K-Ag, K-Pd systems [18]; however, the synthesis of these intermetallic compounds requires pressures of several tens of GPa. Another interesting example is the $CaCo_2$ compound, synthesized under a pressure of 5-7 GPa and possessing fascinating ferromagnetic characteristics such record Curie temperature for the substances of this class $T_c$=528K (Co magnetic moment is 1.75 μB) at normal pressure [60].

An outstanding event in high-pressure chemistry was the synthesis from fullerite $C_{60}$ of a ferromagnetic state of carbon with an extraordinary high Curie temperature $T_C \sim 800$ K [61]. The nature of ferromagnetism in purely carbon materials is not fully understood yet and has long been debated. At the same time, it is hoped that ferromagnetism of carbon materials can also be attained on particular treatment of carbon films at normal pressure.

Recently, new ferromagnetic metals and semiconductors in the systems of tetrahedral semiconductor $A^{III}B^{V}$ (GaSb, GaAs) - 3d-metal of the Fe group (Mn, Cr, Co) have been synthesized under pressure [62]. The above materials are of an immense interest as basic elements of spintronics.

Recently high-pressure synthesis was successfully applied to fabricate new class of materials – multiferroic substances [63] such as $BiMnO_3$, $BiAlO_3$, $BiGaO_3$, $BiFeO_3$ etc. These materials with perovskite-like structure possess simultaneously ferromagnetic and ferroelectric properties (magnetic moment and spontaneous electric polarization).

By affecting the viscosity of materials, high pressures make it possible to obtain metallic glasses Cu-Sn, Cu-Ti, Cu-Zn, Pd-Si, and the like in the bulk form of about several mm in size [19].

Some high-pressure phases, for instance, PbTe with rock-salt structure, exhibit very intriguing thermoelectric properties: the ratio of Seebek coefficient to thermal conductivity several times higher than that of pressure phase [64].

High pressures substantially modify the state diagrams in the systems Al-Si, Al-Ge, expanding the region of the solid solutions Al:Si, Al:Ge in particular [65]. These solid solutions can be conserved in the metastable form at room pressure and have superconductive transition temperature $T_C$ up to 11 K, which is 10 times higher than the $T_C$ values for pure aluminum [65]. The high-pressure synthesis has also permitted obtaining new superconducting materials as silicon-based clathrates $Ba_8Si_{46}$ [66]. High pressures provide the growth of single crystals of mercury-based high temperature superconductors (HTS) with the record values of $T_C \sim 150$ K [67].

An individual mention should be made of the pressure synthesis of a highly boron-doped diamond [68] (Fig.16 a,b). A sufficiently high superconductive transition temperature $T_C \sim 7$ K, huge critical magnetic field $H_{cr} \sim 15$ Tl, and the bulk character of superconductivity give grounds to think of superconducting diamonds as materials of a great promise for the electronics of the future [68].

There are also many examples of new pressure-synthesized dielectric materials. Thus, high-pressure phases of oxides like $SiO_2$ [32], $TiO_2$ [40] are materials of much importance for geophysics and geochemistry. In addition, many high-pressure oxide modifications are superhard materials (see above).

Some dielectric high-pressure phases show interesting optical or dielectric characteristics, for example, the non-linear optical properties in the $CO_2$ high-pressure phase or recovered $SrGeO_3$ perovskites with a high dielectric constant $\varepsilon \sim 400$-$800$ [1].

High pressure synthesis allows to prepare a lot of new oxides with unusual valency of metal ions and new interesting structures [69,70]. Different borates [69] like $DyBO_3$, $Gd_2B_4O_9$, perovskite-structure oxides [70] like $NdNiO_3$ and new ternary dense oxide phases [70] like $Al_3BO_6$ and $AlBO_3$ can be mentioned in this respect.

By using high pressures one can increase the viscosity of definite melts and produce a variety of new insulating and semiconducting glasses, for example $MgF_2 - ScF_3$ [71] or AsS [72].

It is worthy of mention are recent accomplishments in the high-pressure synthesis of different kinds of ice-based clathrates [73]. These materials are of interest as potential accumulators of hydrogen and various hydrocarbons [73]. Among other examples of new high-pressure synthesized low-z element-based compounds, mention should be made of the $P_3N_5$ metastable phase, a phosphorous compound with five-coordinated P-atoms, the first of its kind ever obtained [74]. Of some interest is to probe into the question of the existence of various $P_2O_5$ polymeric and atomic modifications.

More consideration should also be given here to the use of high pressures in the production of gem crystals. However, a hydro-thermal synthesis of emeralds and rubies at moderate pressure falls outside the scope of this review, and the high-pressure synthesis of diamonds has been discussed earlier. Nevertheless, there is one more striking example of how high pressures can be utilized in the manufacture of new materials; namely, for changing the color and transparency of gems, primarily diamonds. Today, modification of the color of brilliants offers almost the only example of an industrial application of high-pressure treatment aside from the already discussed manufacture of superhard materials - diamond and c-BN.

Ideal diamond crystals must be absolutely transparent for visible spectrum. However, the overwhelming number of natural diamonds has color that is more rarely a tint of yellow or blue and more often an unattractive brown tint due to impurity centers associated with nitrogen or boron impurities and due to plastic defects.

High pressure – high temperature treatment is a means of refining the color of diamonds. In the process, some defects are eliminated and the coloring centers are modified. Colorless or light-brown nitrogen-containing diamonds of Ia type can be transformed into bright yellow, green or red diamonds [75] (Fig.17a,b), and brown nitrogen-free natural diamonds of IIa type are turned into colorless ones [76] (Fig.18). The maximum weight of work diamond crystals is dictated by the possibilities of a high-pressure apparatus and currently amounts to 20-30 carat per experiment (Fig.18b). In this case, the market value of diamonds after treatment can grow by dozens of times, and diamond coloration or decoloration commercial applications are expected to be vigorously developing in the coming years.

## 7. Molecular Compounds and New Polymers: Transfiguration under Pressure

In this section, we get back to the high-pressure treatment of molecular substances and to the production of new materials, among them polymers. In so doing, we will deal with hypothetical new materials rather than with the examples of already existing ones. As was noted in Sections 3,4, by subjecting molecular substances to the high-pressure treatment one can often obtain new substances with a high temperature stability. This can be achieved in the event that pristine molecular substances themselves are not in the most equilibrium state. In Section 4, production of polyethylene from ethylene was cited as an example. Similarly, acetylene $C_2H_2$, benzene $C_6H_6$ and other hydrocarbons under pressure irreversibly transform to polymeric modifications and then, at higher pressures, to amorphous diamond-like carbon saturated by hydrogen a-C:H [77]. The polymerization process is often made easier if simultaneously employing high pressure and laser photo-excitation [78].

In the N-H system at normal pressure there is only one equilibrium phase – ammonia $NH_3$, while other molecular compounds are metastable. In the C-N system, also metastable with regard to the decomposition to solid carbon and molecular nitrogen are molecular phases CN, NCN, CNC, CCN, $C_2N_2$, $C_3N_4$ and the like [25]. All these metastable compounds under pressure also can irreversibly transform into new phases, including polymeric ones. Carbon-containing polymeric phases can have high temperature stability at normal pressure. Initially metastable phases are many molecular compounds comprising atoms of phosphorous and sulfur. Under pressure, they also can turn into polymeric modifications with high temperature stability. Here we again point out to the irreversibility of the transformation of metastable molecular phases into polymeric or atomic modifications; on annealing, such "high-pressure" phases never transform back to the initial molecular states.

Let us discuss such substance as molecular carbon oxide in greater detail. In the carbon-oxygen system at normal pressure there is only one stable compound – $CO_2$ (the $CO_2$ formation energy is 1598 kJ/mole, CO – 1071 kJ/mole [25]). The CO molecular condensed phases are metastable with relation to the decay into the mixture of solid carbon (graphite) and $CO_2$; this distinguishes the CO compound from its isoelectronic analog $N_2$, for which the molecular phases at normal pressure are equilibrium. The study of the CO phase diagram at high pressures has revealed the transformation to a CO polymerized modification, erroneously taken to be a high-pressure modification of the molecular phase[79]. In actual fact the polymerization of the CO molecular phase is a non-equilibrium kinetic transformation to a more low-laying energy state. Indeed, according to the computer simulation data, the CO polymerization is followed by a significant heat release [80]. The CO polymerized phase is metastable at room pressure up to high temperatures (400-700 K) [79], which is much higher than the melting temperature of a CO molecular crystal. If heating the CO polymeric phase at normal pressure, a significant amount of heat is being released, which was mistakenly ascribed in [79] to the energy stored up in the polymerized phase at the expense of the PV-term in the Gibbs' free energy. Meanwhile, a simple calculation of the difference of the PV-terms for molecular and polymer modifications gives the value ~ 0.5 - 1 $kJg^{-1}$, which is an order of magnitude less than the observed ones. The observed energy release, indeed, is associated with the transition of CO-polymer at heating to the equilibrium mixture of $CO_2$ and solid carbon (these are precisely the products found after the annealing of the CO polymer). Note that the pattern of the high-pressure transformation of the CO molecular phase can be more complicated; in particular, a disproportination is possible of the CO compound into $C_2O_3$ and $C_3O_2$ modifications which, in turn, are polymerized [79,80]. Thus, the polymerization of CO under pressure is a non-equilibrium transformation to a more low-laying energy state, similar to the polymerization of acetylene $C_2H_2$, ethylene $C_2H_4$ and benzene $C_6H_6$ under pressures. High pressure in this case is a factor for lowering the energy barrier for the transformation of the metastable molecular phases to more equilibrium states.

At the same time, at sufficiently high pressures the equilibrium concentration diagrams themselves can change; specifically, a number of saturated hydrocarbons can become more advantageous in terms of energy than the mixture of condensed phases of methane and carbon. As a result, considerably complex polymerized phases of low-Z elements may emerge at compression not only as energy-intermediate states, but also as high-pressure ground-state modifications. For example, at certain pressures the formation of complex hydrocarbons becomes beneficial, which is of great importance for the abiogenic synthesis of petroleum (petroleum at ambient pressure is, of course, composed of the metastable phases of hydrocarbons) [81].

To sum up, all P,T-diagrams of the light element-based compounds can be divided into three zones (Fig.19).

Zone I includes moderate pressures and temperatures (~1-5 GPa, 10 -1000 K), at which the Gibbs' free energy of the phases is almost completely determined by the bonding energy of atoms in molecules. Most of molecular substances in this zone are in the metastable state.

Zone III covers ultrahigh pressures (~20-50 GPa), at which the Gibbs' free energy of the majority of the phases is largely determined by the PV-term. The substances in this zone are in the stable state, as a rule, and represent the mixture of simple solid equilibrium modifications (diamond, $CO_2$, $H_2O$, etc.). Specifically, at megabar pressures P ~ 20-50 GPa and moderate temperatures the PV-contribution to the Gibbs' free energy facilitates the transition of all unsaturated hydrocarbons to a solid hydrogen solution in diamond; at higher temperatures it gives rise to the formation of a diamond-hydrogen mixture. As was pointed out above, the formation of amorphous diamond-like carbon saturated with hydrogen a-C:H, indeed, was observed for many hydrocarbons at ultrahigh pressures [77].

Zone II corresponds to intermediate pressures P~ 1-20 GPa and temperatures 200-1500 K, at which the formation of a large number of both "kinetic" and equilibrium, at a given pressure, polymerized phases is possible. The contributions to the Gibbs' free energy from intra-molecular, intermolecular interaction and the PV-contribution for these phases are comparable in magnitude. One should only remember that the modifications in question, as a rule, are not equilibrium high-pressure phases of molecular substances, since the majority of molecular substances at ambient pressure are metastable themselves. The high-pressure experiments in this case again send a "prompting message" with regard to the possibility of the existence of new energy-intermediate polymer phases. Evidently, many of these uncommon polymers can be synthesized through "chemical" techniques at normal pressure.

Finally, let me note that the problem of the abiogenic synthesis of petroleum and other organic compounds is closely linked to the problem of the origin of life [82]. However, since this subject is very extensive, important and controversial, it will not be dealt with in the present review devoted to more or less simple high-pressure synthesized materials.

## Conclusion

In summary it must be emphasized that relatively moderate pressures ~ 0.1 GPa are extremely intensively used industrially. In addition, there is a great diversity of examples of the pressure effect on the phase composition and microstructure of the substances in the pressure range 1 – 100 GPa. However, the commercial production of these new materials is severely restricted because of the small volume of an obtained product at such pressures. As a result, high pressures on an industrial scale, aside from the synthesis of superhard materials proper, are only used for modifying the color and transparency of brilliants. All new promising semiconductors, superconductors, magnets, obtained in abundance by means of high pressure, still remain within the scope of laboratory research, although the very existence of these materials gives production engineers a hint at the possibility of obtaining them through other methods. This, even to a greater extent, refers to new phases, among them polymeric ones, originated from low-Z element molecular compounds. Many new polymers in the C-N-O-H systems can apparently be obtained without the application of pressure by means of photo-chemistry, catalysts, and through other chemical methods. This area of helpful hints from the science of high-pressure materials is not yet appreciated at its true value, and we expect the next decades to abound with discoveries of amazing new substances.

## Acknowledgements

The author is grateful to A.G. Lyapin, S.V. Popova, S.M. Stishov, N.A. Nikolayev, F.S. Yel'kin, V.L. Solozhenko, N.A. Bendeliani, A.G. Sinyakov, and L.B. Solodukhina for the assistance and useful discussions. The work was supported by the Russian Foundation for Basic Research (projects nos. 05-02-16596 and 04-02-16308), the Program of the Presidium of the Russian Academy of Sciences, and the Russian Science Support Foundation.

**SUBSRIPTIONS TO THE FIGURES**

Fig.1  Schematic P,T-map of the scientific domains

Fig.2 Main types of high-pressure devices
- a) Internationally used high-pressure apparatus: (1) piston-cylinder; (2) Bridgeman anvils; (3) belt; (4) multianvil
- b) High-pressure apparatus designed in Russia: (1) Chechevitsa; (2) Toroid
- c) Toroid anvils with the high-pressure cells after 10 GPa pressure treatment
- d) 50 000 ton ultralarge-volume press operating in the Institute for High Pressure Physics (IHPP RAS), Russia

Fig.3 Energy minima in the configuration space. At pressures and temperatures below critical parameters metastable "kinetic" phases can transform into one another but can not transit to the ground stable state with the minimal Gibbs' free energy. At critical P,T-parameters the irreversible transitions from the "kinetic" phases to the stable modification occur for the experimental times

Fig.4 Solid-solid transformation from high - to low-pressure phases. During pressure decrease the transformation of high-pressure phase II to the low-pressure phase I takes place not at the equilibrium line (solid line), but at some kinetic line (curves (a) and (b)). Below this kinetic curve, the high-pressure phase II can exist in metastable form for a long time. In the case of (a) the high-pressure phase can't be conserved at normal pressure at any temperatures. Point (1) corresponds to the metastable phase conserved at room pressure for the case (b)

Fig.5 Gibbs' free energies vs temperature dependencies for stable and metastable phases. Phase relations between crystalline and liquid phases as well as melting temperature $T_m$ are defined by the Gibbs' energy , that must be minimal for the stable phase. If  the substance has an additional metastable solid phase, the temperature of equilibrium between the metastable solid and stable liquid $T_m^*$ is always below $T_m$

Fig.6 Diagram of possible ways of producing new materials by high-pressure methods

Fig.7 Different grains morphology of $GeO_2$ high-pressure phases obtained by using different P,T-paths onto phase diagram.
- a) At low pressures - high temperatures conditions of treatment a round grain morphology of $GeO_2$ high-pressure phases is observed
- b) At high pressures - low temperatures conditions of treatment a lamel grain morphology of $GeO_2$ high-pressure phases is observed

Fig.8 Transitional phase diagram of initially metastable phases. The grey band corresponds to the irreversible non-equilibrium relaxation of metastable phases to the stable modifications and separates the quasi-equilibrium and equilibrium P,T-regions. Inside these regions the transformations (1-2, 2-3, 1-3 and 4-5) are reversible. The transitions 3-4 and 3-5 are irreversible

Fig.9 Diagram illustrating the classes of new high-pressure-synthesised materials

Fig.10 Hardness vs shear, Young's and bulk moduli dependencies for superhard materials according to the data collected in [26]. Dashed lines illustrate the correlations between hardness and shear or Young's moduli

Fig. 11 Correlations between the moduli (bulk and shear) and effective electron valence density.
   a) Dependence of the bulk modulus of solid elemental substances on effective electron valence density
   b) Dependence of the shear modulus of solid elemental substances on effective electron valence density (upper figure) and the same dependence for elements from the definite subgroups (bottom figure)

Fig. 12 The 2-4mm-sized single crystals of stishovite prepared in IHPP RAS

Fig. 13 Transitional P,T-phase diagram of $C_{60}$ fullerite. The corresponding phases are designated as follows: fcc, sc, and gc correspond to monomeric phases; O1, O2, R, and T are 1-d orthorhombic phases, and 2d-rhombohedral and tetrahedral phases; $d-C_{60}$ is a phase with dimmers; $nc-sp^2$- is a disordered graphite-like phase; 3D is a 3D polymer; $a-sp^3$ and $a(nc)-sp^2-sp^3$ are amorphous and nanocrystalline phases (see [24] for the details)

Fig. 14 Mechanical properties vs density of carbon phases
   a) Hardness vs density for different carbon phases obtained from $C_{60}$ (see [24] for the details)
   b) Bulk modulus vs density for carbon modifications obtained from $C_{60}$ and theoretical data for hypothetical carbon phases (see [24] for the details)

Fig.15 A single crystal of the $TiO_2$ high-pressure phase with the $PbO_2$-structure type (a) and corresponding indexing of the crystal faces (b)

Fig. 16 Superconducting diamond
   a) a polycrystalline boron-doped diamond prepared by high pressure- high temperature treatment
   b) the temperature dependence of the electrical resistance of the sample in the vicinity of the superconducting temperature and temperature dependence of the superconducting gap

Fig. 17 Examples of the coloration of natural diamonds by high pressure- high temperature treatment
   a) light-brown nitrogen-containing diamonds transformed into green diamonds
   b) bright yellow diamonds prepared from very light-brown diamonds

Fig. 18 Examples of the decoloration of brown nitrogen-free natural diamonds by high pressure- high temperature treatment
   a) diamonds before and after high pressure- high temperature treatment
   b) record sized 21 carat decolorated diamond before and after high pressure- high temperature treatment

Fig.19 General kind of a transitional P,T-phase diagram of molecular low Z- elements compounds. In most cases the transitions between the phases from Zone I to Zone II and from Zone II to Zone III are non-equilibrium "kinetic" transformations.

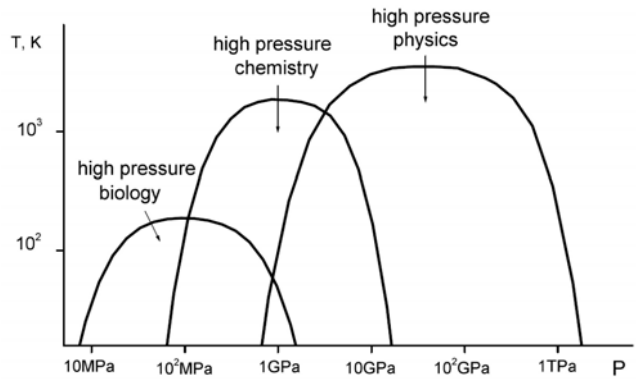

Fig. 1. Schematic P,T-map of the scientific domains.

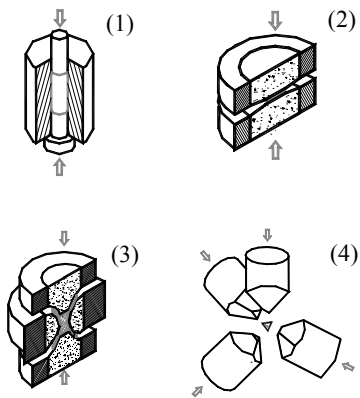

Fig. 2a.

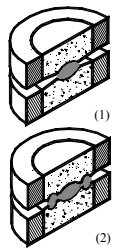

Fig. 2b.

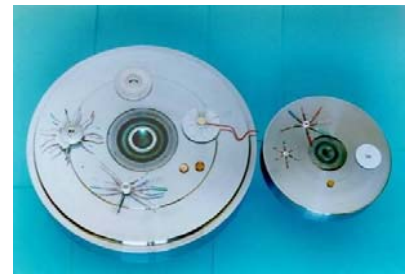

Fig. 2c.

Fig. 2d.

Fig. 2d Main types of high-pressure devices:
   a) Internationally used high-pressure apparatus: (1) piston-cylinder; (2) Bridgeman anvils; (3) belt; (4) multianvil
   b) High-pressure apparatus designed in Russia: (1) Chechevitsa; (2) Toroid
   c) Toroid anvils with the high-pressure cells after 10 GPa pressure treatment
   d) 50 000 ton ultralarge-volume press operating in the Institute for High Pressure Physics (IHPP RAS), Russia.

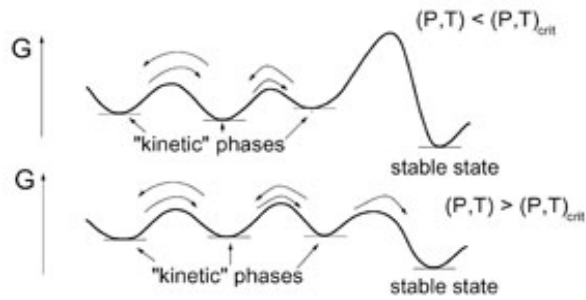

Fig. 3. Energy minima in the configuration space. At pressures and temperatures below critical parameters metastable "kinetic" phases can transform into one another but can not transit to the ground stable state with the minimal Gibbs' free energy. At critical P,T-parameters the irreversible transitions from the "kinetic" phases to the stable modification occur for the experimental times

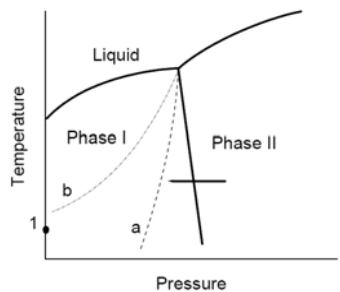

Fig. 4. Solid-solid transformation from high - to low-pressure phases. During pressure decrease the transformation of high-pressure phase II to the low-pressure phase I takes place not at the equilibrium line (solid line), but at some kinetic line (curves (a) and (b)). Below this kinetic curve, the high-pressure phase II can exist in metastable form for a long time. In the case of (a) the high-pressure phase can't be conserved at normal pressure at any temperatures. Point (1) corresponds to the metastable phase conserved at room pressure for the case (b).

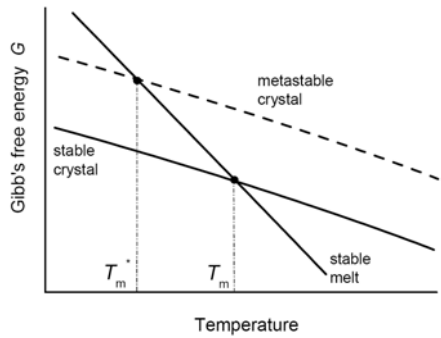

Fig. 5. Gibbs' free energies vs temperature dependencies for stable and metastable phases. Phase relations between crystalline and liquid phases as well as melting temperature $T_m$ are defined by the Gibbs' energy , that must be minimal for the stable phase. If the substance has an additional metastable solid phase, the temperature of equilibrium between the metastable solid and stable liquid $T_m^*$ is always below $T_m$.

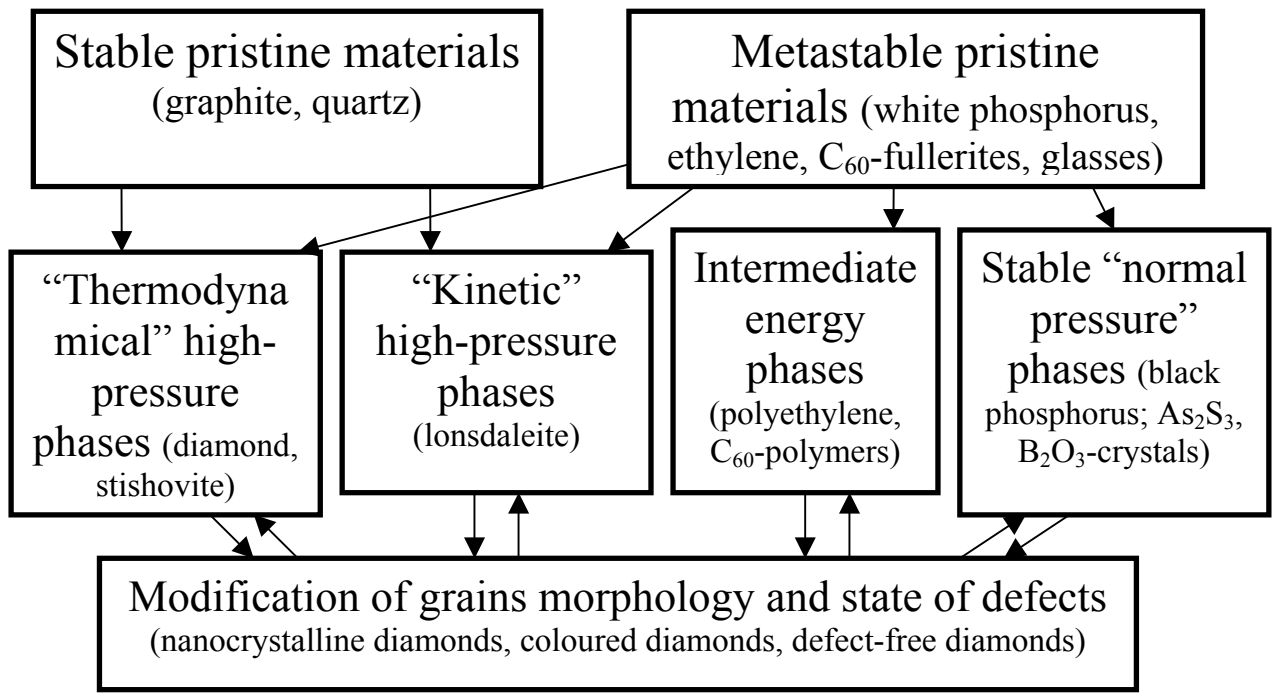

Fig. 6. Diagram of possible ways of producing new materials by high-pressure methods.

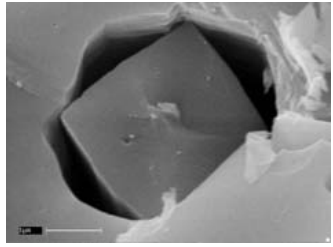

a)

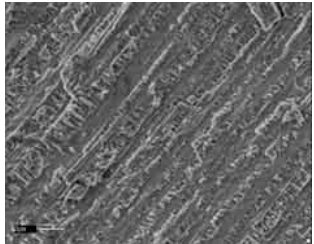

b)

Fig. 7. Different grains morphology of $GeO_2$ high-pressure phases obtained by using different P,T-paths onto phase diagram.
- a) At low pressures - high temperatures conditions of treatment a round grain morphology of $GeO_2$ high-pressure phases is observed.
- b) At high pressures - low temperatures conditions of treatment a lamel grain morphology of $GeO_2$ high-pressure phases is observed.

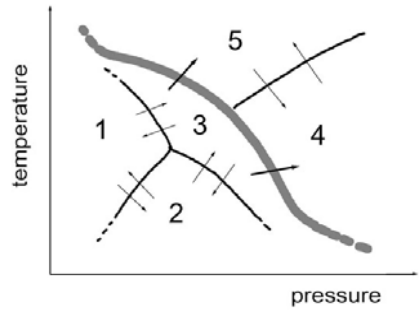

Fig. 8. Transitional phase diagram of initially metastable phases. The grey band corresponds to the irreversible non-equilibrium relaxation of metastable phases to the stable modifications and separates the quasi-equilibrium and equilibrium P,T-regions. Inside these regions the transformations (1-2, 2-3, 1-3 and 4-5) are reversible. The transitions 3-4 and 3-5 are irreversible.

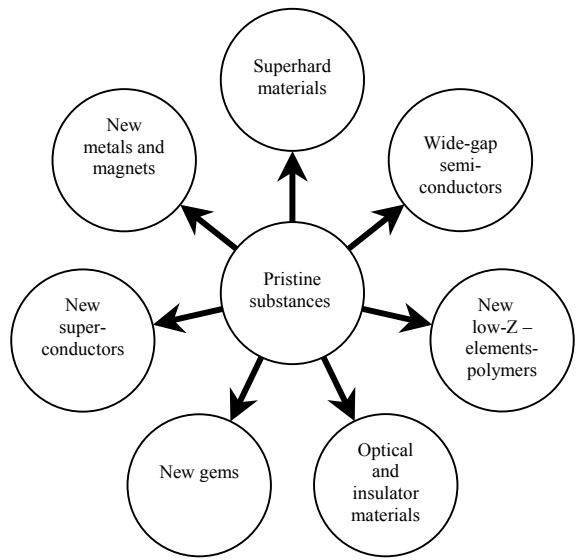

Fig. 9. Diagram illustrating the classes of new high-pressure-synthesised materials.

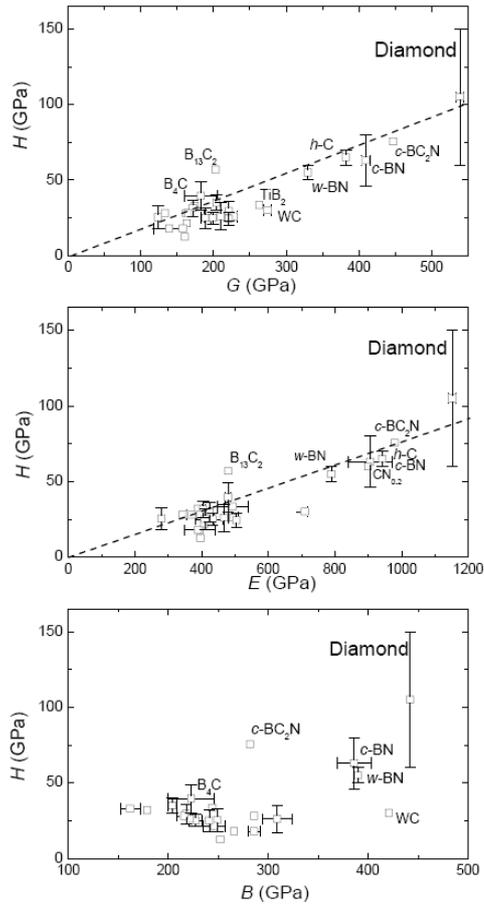

Fig. 10. Hardness vs shear, Young's and bulk moduli dependencies for superhard materials according to the data collected in [26]. Dashed lines illustrate the correlations between hardness and shear or Young's moduli.

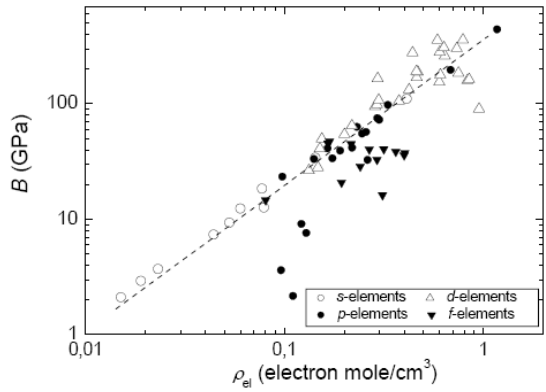

Fig. 11a.

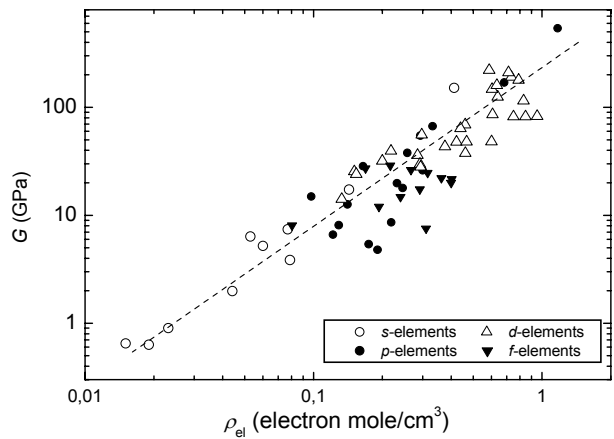

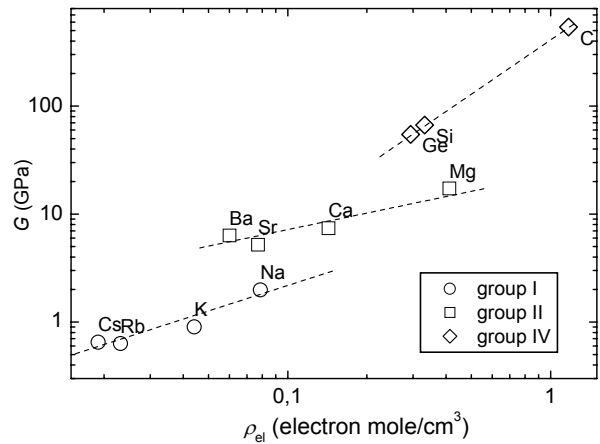

b)

Fig. 11. Correlations between the moduli (bulk and shear) and effective electron valence density.
a) Dependence of the bulk modulus of solid elemental substances on effective electron valence density/
b) Dependence of the shear modulus of solid elemental substances on effective electron valence density (upper figure) and the same dependence for elements from the definite subgroups (bottom figure).

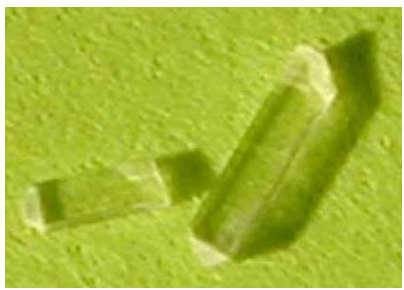

Fig. 12. The 2-4mm-sized single crystals of stishovite prepared in IHPP RAS.

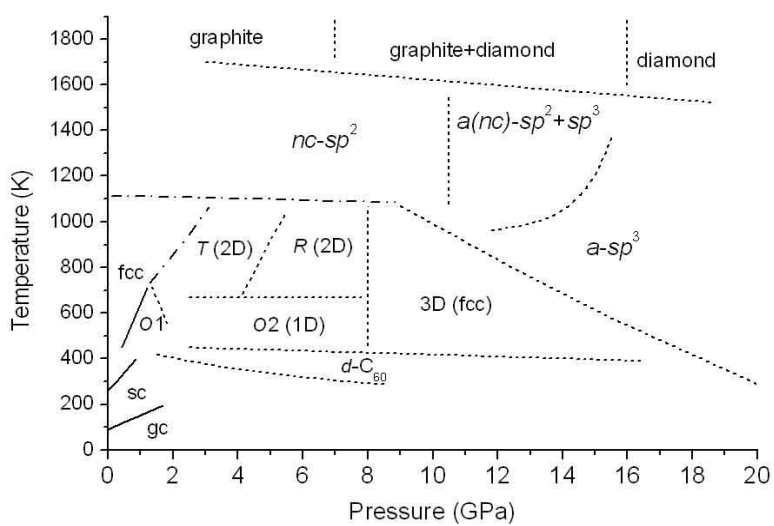

Fig. 13. Transitional P,T-phase diagram of $C_{60}$ fullerite. The corresponding phases are designated as follows: fcc, sc, and gc correspond to monomeric phases; O1, O2, R, and T are 1-d orthorhombic phases, and 2d-rhombohedral and tetrahedral phases; d-$C_{60}$ is a phase with dimmers; nc-$sp^2$- is a disordered graphite-like phase; 3D is a 3D polymer; a-$sp^3$ and a(nc)-$sp^2$-$sp^3$ are amorphous and nanocrystalline phases (see [24] for the details).

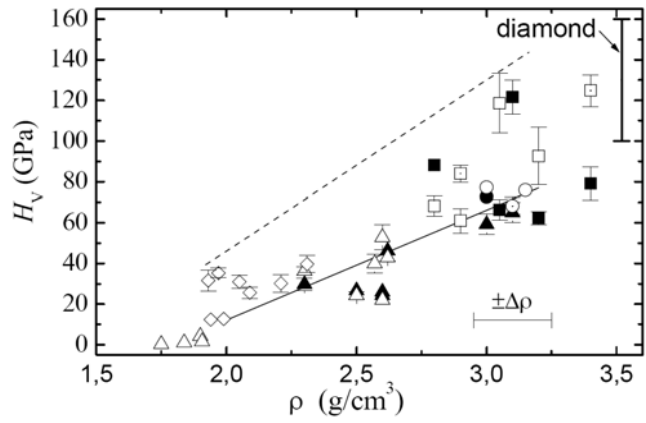

Fig. 14a.

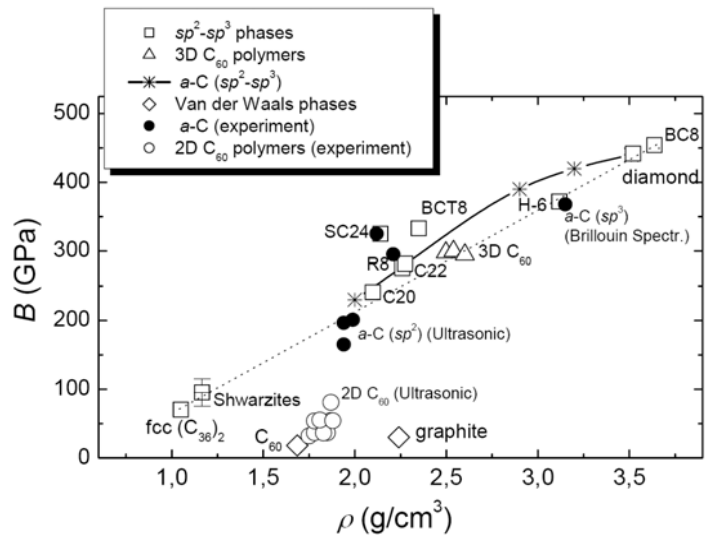

Fig. 14b.

Fig. 14 Mechanical properties vs density of carbon phases.
 a) Hardness vs density for different carbon phases obtained from $C_{60}$ (see [24] for the details)
 b) Bulk modulus vs density for carbon modifications obtained from $C_{60}$ and theoretical data for hypothetical carbon phases (see [24] for the details)

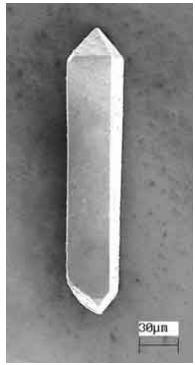

a)

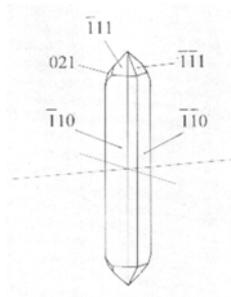

b)

Fig.15. A single crystal of the TiO$_2$ high-pressure phase with the PbO$_2$-structure type (a) and corresponding indexing of the crystal faces (b).

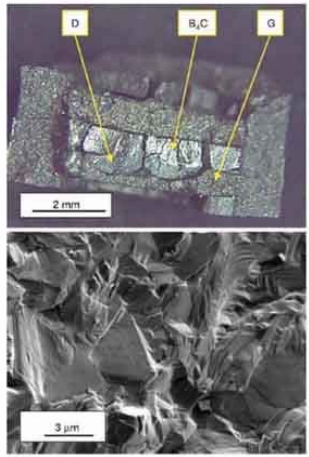

Fig. 16a.

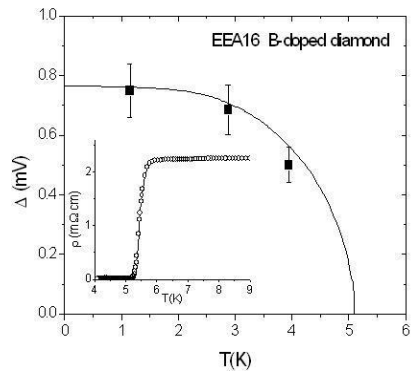

Fig. 16b.

Fig. 16 Superconducting diamond
  a) a polycrystalline boron-doped diamond prepared by high pressure- high temperature treatment.
  b) the temperature dependence of the electrical resistance of the sample in the vicinity of the superconducting temperature and temperature dependence of the superconducting gap

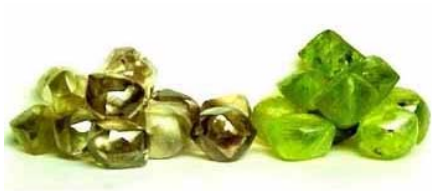

a)

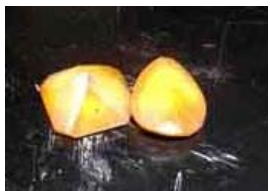

b)

Fig. 17. Examples of the coloration of natural diamonds by high pressure- high temperature treatment
   a) light-brown nitrogen-containing diamonds transformed into green diamonds
   b) bright yellow diamonds prepared from very light-brown diamonds.

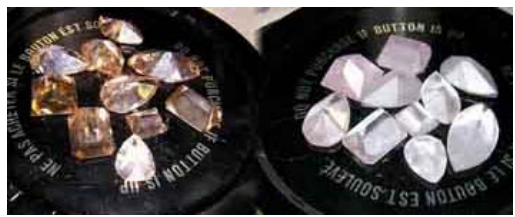

a)

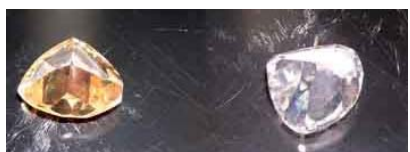

b)

Fig. 18. Examples of the decoloration of brown nitrogen-free natural diamonds by high pressure-high temperature treatment
   a) diamonds before and after high pressure- high temperature treatment
   b) record sized 21 carat decolorated diamond before and after high pressure- high temperature treatment

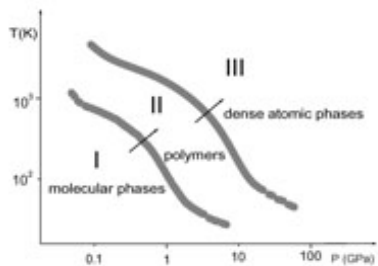

Fig. 19. General kind of a transitional P,T-phase diagram of molecular low Z-elements compounds. In most cases the transitions between the phases from Zone I to Zone II and from Zone II to Zone III are non-equilibrium "kinetic" transformations.